\documentclass[11pt]{article}
\usepackage{amssymb,amsmath,cite,epsfig}
\usepackage[a4paper,top=2.5cm, bottom=2.5cm, left=2.5cm, right=2.5cm]{geometry}

\begin{document}

\begin{titlepage}
\vspace*{3cm}
\begin{center}
{\Large \textsf{\textbf{Non-commutative DKP equation and pair creation\\in curved space-time}}}
\end{center}
\vskip 5mm
\begin{center}
{\large \textsf{Lamine Khodja$^{1}$ and Slimane Zaim$^{2}$}}\\
\vskip 5mm
$^{1}$D\'{e}partement de Physique, Facult\'{e} des Math\'{e}matiques et des Sciences de la Mati\'{e}re\\
Universit\'{e} Kasdi Merbah - Ouargla, Algeria.\\
$^{2}$D\'{e}partement de Physique, Facult\'{e} des Sciences  de la Mati\`{e}re\\
Universit\'{e} Hadj Lakhdar -- Batna 1, Algeria.\\
\end{center}
\vskip 2mm
\begin{center}{\large\textsf{\textbf{Abstract}}}\end{center}
\begin{quote}
This study is about the application of the non-commutativity on the DKP equation up to first order in $\theta$ for the process of pair-creation of bosons from vacuum in (1+1) curved space-time. The density of particles created in the vacuum can be calculated with the help of the Bogoliubov transformations. The non-commutative density of created particles is found to decrease as $1/\sqrt{\theta}\sim\Lambda_{\mathrm{NC}}$, so that the rate of particle creation increases whenever a non-commutativity parameter is small and this corresponds to the spirit of quantum mechanics.
\end{quote}
\vspace*{2mm}
%
\end{titlepage}

\section{Introduction}

Several studies of Duffin-Kemmer-Petiau (DKP) theory in pure Riemann-Cartan space have been considered \cite{1a,2a,3a,4a}, and pair production problems for spin-0, spin-1/2 and spin-1 massive particles in curved space-time in the presence of an external electromagnetic field have been analyzed \cite{5a,6a,7a,8a,9a,10a,11a,12a,13a,14a}.

The particle creation mechanism in curved space is an important concept for problem in the vicinity of black holes as well as in expanding cosmological universes. The curvature of space-time affects the excitation of the gravitation field. As time elapses the curvature could excite the field, which is related to the number of particles in the system. The curvature of space gives rise to the important property of torsion in curved space which itself leads to non-commutative coordinates of the point between them.

Here we point out that the concepts of classical geometry may not be quite appropriate to describe the physical phenomenon at short distances. For example the standard concept of space-time as a geometric manifold should break down at very short distances of the order of the planck length. This led physicists to speculate that space-time becomes non-commutative at very short distances when quantum gravity becomes relevant. Therefore several solutions were suggested to understand this problem, including the non-commutativity as a mathematical concept expressing uncertainty in quantum mechanics, where it applies to any pair of conjugate variables such as position and momentum.

It would be very interesting to look for a mechanism of creating particles in the non-commutative geometry. Moreover, in string theories, the non-commutative gauge theory appears as a certain limit in the presence of a background field. In this context, a gauge field theory with star products and Seiberg-Witten maps is used \cite{15a,16a,17a}.

The purpose of this article is to study the particle production mechanism based on the non-commutative DKP equation of massive particles by considering the conjugate coupling of gravitation and non-commutativity of curved space-time. Our goal is to calculate the number density of the particles created under the effect of non-commutativity and in the presence of a gravitational field using vacuum mode solutions. We show how the non-commutativity effects influence particle creation. The important new result in this work is that non-commutativity effects induce particle creation more effectively than a gravitation field.

This paper is organized as follows. In the first part we present the relevant relations in the non-commutative space-time and derive the corresponding Seiberg-Witten maps up to first order in $\theta$. In the second part we present the modified DKP equation and compute the number density of the created particles in curved space-time. Finally, we draw our conclusions.

\section{Formalism and non-commutative DKP equation}

The non-commutative space-time is characterized by the operators $\widehat{x}^\mu$ satisfying the canonical-type relation:
\begin{equation}
\left[\widehat{x}^\mu,\widehat{x}^\nu\right]_\ast=i\,\theta^{\mu\nu}\,,
\end{equation}
where the star Moyal product $\ast$ is defined between two fields $\phi(x)$ and $\psi(x)$ by:
\begin{equation}
\phi(x)\ast\psi(x)=\exp\left(\frac{i}{2}\,\theta^{\mu\nu}\,\left.\frac{\partial}{\partial x^\mu}\,\frac{\partial}{\partial y^\nu}\,\phi(x)\,\psi(y)\right\vert_{y=x}\right),
\end{equation}
where $\theta^{\mu\nu}$ is a constant real matrix.

One can get at first order in the non-commutative parameter the following Seiberg-Witten maps \cite{14a,15a,16a,17a,18a,19a}:
\begin{align}
\widehat{\psi}&=\widetilde{\psi}+\frac{1}{4}\,\theta^{\mu\nu}\,A_\nu\,\partial_\mu\,\widetilde{\psi}+\frac{i}{8}\,\theta^{\mu\nu}\left[A_\mu,A_\nu\right]\widetilde{\psi}\,,\\
\widetilde{\psi}&=\psi+\frac{1}{4}\,\theta^{\mu\nu}\,\Gamma_\nu\,\partial_\mu\,\psi+\frac{i}{8}\,\theta^{\mu\nu}\left[\Gamma_\mu,\Gamma_\nu\right]\psi\,,\\
\widehat{A}_\mu &=A_\mu+\frac{1}{4}\,\theta^{\rho\sigma}\left\{A_\sigma,\partial_\rho\,A_\mu\right\}+\frac{1}{4}\,\theta^{\rho\sigma}\left\{F_{\rho\mu},A_{\sigma}\right\},\\
\widehat{\omega}_\mu^{ab}&=\omega_\mu^{ab}-\frac{1}{2}\,\theta^{\rho\sigma}\left\{\omega_\rho,\partial_\sigma\,\omega_\mu+R_{\sigma\mu}\right\}^{ab}\,,
\end{align}
where $\psi$, $A_\mu$ and $\omega_\mu^{ab}$ are the Duffin-Kemmer-Petiau spinor, gauge and spin connection respectively. $F_{\rho\mu}$ is the field strength tensor with $A_\mu$ and $R_{\sigma\mu}^{ab}$ representing the curvature of Riemann-Cartan space-time.

We propose the following non-commutative action in a curved non-commutative space-time:
\begin{equation}
\mathcal{S}=\frac{1}{2\,\kappa^2}\int \mathrm{d}^dx\left(\mathcal{L}_{\mathrm{DKP}}+\mathcal{L}_\mathrm{G}\right),
\end{equation}
where $\mathcal{L}_\mathrm{G}$ and $\mathcal{L}_{\mathrm{DKP}}$ stand for the pure gravity and matter scalar densities corresponding to the charged vector bosons in the presence of an electric field:
\begin{align}
\mathcal{L}_\mathrm{G} &=\sqrt{-g}\ast\widehat{R}\,,\\
\mathcal{L}_{\mathrm{DKP}}&=\sqrt{-g}\ast\left[\,\overline{\widehat{\psi}}\ast\left(i\,\beta^\mu\ast\partial_\mu\,\widehat{\psi}-i\,\beta^\mu\ast\Gamma_\mu\ast\widehat{\psi}\right)-m\,\overline{\widehat{\psi}}\ast\widehat{\psi}\right],
\end{align}
with, in flat space:
\begin{align}
\Gamma_\mu&=\omega_\mu^{ab}\,S_{ab}\,,&S_{ab}&=\left[\beta_a,\beta_b\right],\\
\overline{\widehat{\psi}}&=\widehat{\psi}^\dag\,\eta^0\,,&\eta^a&=2\left(\beta^a\right)^2-\eta^{aa}\,,
\end{align}
with $\beta^a$ being the DKP matrices which satisfy the following trilinear commutation relation:
\begin{equation}
\beta^a\,\beta^b\,\beta^c+\beta^c\,\beta^b\,\beta^a=\beta^a\,\delta^{bc}+\beta^{c}\,\delta^{ba}\,.
\end{equation}

Then using the principle of least action one can deduce the Euler-Lagrange equations \cite{16a, 17a}:
\begin{equation}
\frac{\partial \mathcal{L}}{\partial\widehat{\psi}}-\partial_\mu\left(\frac{\partial\mathcal{L}}{\partial\left(\partial_\mu\widehat{\psi}\right)}\right)+ \partial_\mu\,\partial_\nu\left(\frac{\partial\mathcal{L}}{\partial\left(\partial_\mu\,\partial_\nu\,\widehat{\psi}\right)}\right)=0\,.\label{eq}
\end{equation}
Using the modified field equation \eqref{eq} one can find, in a free non-commutative curved space-time, the following modified DKP equation:
\begin{multline}\label{eq:fourteen}
\left(i\,\beta^\mu\left(\partial_\mu-\Gamma_\mu\right)-m\right)\psi-\frac{1}{2}\,\theta^{\alpha\beta}\left[\left(\partial_\alpha\,\beta^\mu\right)\left(\partial_\mu\,\partial_\beta\,\psi\right)-\partial_\alpha\left(\beta^\mu\, \Gamma_\mu\right)\left(\partial_\beta\,\psi\right)\right]-\\
-\frac{i}{2}\,\theta^{\alpha\beta}\,\partial_\alpha\left(\ln\sqrt{-g}\right)\partial_\beta\left[\left(i\,\beta^\mu\left(\partial_\mu-\Gamma_\mu\right)-m\right)\psi\right]=0\,.
\end{multline}

In what follows, we take $x^0=t$ and $x^1=x$. Let us consider the (1+1)-dimensional de-Sitter universe described by the line element:
\begin{equation}
ds^2=-dt^2+e^{2\,H\,t}\,dx^2\,.\label{a}
\end{equation}
In order to simplify the calculations, we take the following components of the dimensionless non-commutative parameter $\theta^{\alpha\beta}$ as:
\begin{equation}
\theta^{\alpha\beta}=\left(\begin{array}{rr}
0&\theta\\-\theta&0\end{array}\right), \qquad \alpha,\,\beta=0,\,1\,.
\end{equation}
with $\theta$ being a real positive constant and the non-vanishing components of the corresponding metric are:
\begin{equation}
g_{\alpha\beta}=\left(\begin{array}{cc}
-1 & 0 \\0 & e^{2Ht}\end{array}\right),\qquad \widetilde{\partial}_0\left(\ln\sqrt{-g}\right)=H\,.
\end{equation}
The symbol ``$\widetilde{\phantom{\partial}}$'' denotes the coordinates in curved space. In this universe, the modified non-commutative DKP equation \eqref{a} takes the form:
\begin{multline}\label{G}
\left(i\,\beta^0\,\widetilde{\partial}_0+i\,e^{-H\,t}\,\beta^1\,\widetilde{\partial}_1-i\,\beta^0\,\widetilde{\Gamma}_0-i\,e^{-H\,t}\,\beta^1\,\widetilde{\Gamma}_1-m\right)\psi-\\
-\frac{1}{2}\,\theta\left[-H\,\beta^1\,\widetilde{\partial}_1^2\,\psi-\beta^0\left(\widetilde{\partial}_0\,\widetilde{\Gamma}_0\right)\left(\widetilde{\partial}_1\,\psi\right)+H\,e^{-H\,t}\,\beta^1\,\widetilde{\Gamma}_1
\left(\widetilde{\partial}_1\,\psi\right)-e^{-H\,t}\,\beta^1\left(\widetilde{\partial}_0\,\widetilde{\Gamma}_1\right)\left(\widetilde{\partial}_1\,\psi\right)\right]-\\
-\frac{i}{2}\,\theta\,H\left(i\,\beta^0\,\widetilde{\partial}_0+i\,e^{-H\,t}\,\beta^1\,\widetilde{\partial}_1-i\,\beta^0\,\widetilde{\Gamma}_0-i\,e^{-H\,t}\,\beta^1\,\widetilde{\Gamma}_1-m\right)\widetilde{\partial}_1\,\psi=0\,,
\end{multline}
with $\beta^0$ and $\beta^1$ being the beta matrices in flat space. Working with the conformal time defined by:
\begin{equation}
\eta=-\frac{1}{H}\,e^{-H\,t}\,,\qquad \widetilde{\partial}_{0}=-H\,\eta\,\partial_\eta\,,
\end{equation}
we write the equation \eqref{G} as:
\begin{multline}
\left(-i\,\beta^0\,H\,\eta\,\partial_\eta-\beta^1\,H\,\eta\,\widetilde{\partial}_1+H\,\eta\,\beta^1\,\widetilde{\Gamma}_1-m\right)\psi-\\
-\frac{1}{2}\,\theta\left[-H\,\beta^1\,\widetilde{\partial}_1^2\,\psi-H^2\,\eta\,\beta^1\,
\widetilde{\Gamma}_1\left(\widetilde{\partial}_1\,\psi\right)-H^2\,\eta^2\,\beta^1\left(\partial_\eta\,\widetilde{\Gamma}_1\right)\left(\widetilde{\partial}_1\,\psi\right)\right]-\\
-\frac{i}{2}\,\theta\,H\left(-i\,\beta^0\,H\,\eta\,\partial_\eta-\beta^1\,H\,\eta\,\widetilde{\partial}_1+H\,\eta\,\beta^1\,\widetilde{\Gamma}_1-m\right)\widetilde{\partial}_1\,\psi=0\,.
\end{multline}

In 2D space-time, we can write the spin connections as:
\begin{align}
\widetilde{\Gamma}_1&=-\frac{1}{2\,\eta}\left(\alpha^1\otimes I+I\otimes\alpha^1\right),\\
\alpha^1&=\gamma^0\,\gamma^1\,,\qquad \textrm{with}\qquad \left(\gamma^0,\gamma^1\right)=\left(\sigma^3,-i\,\sigma^2\right),
\end{align}
where $\otimes$ describes the Kronecker product of two matrices $A$ and $B$ \cite{cronecker}:
\begin{equation}
A\otimes B=\left(\begin{array}{ccc}
a_{11}\,B & \cdots & a_{1n}\,B\\
\vdots & \ddots & \vdots \\
a_{m1}\,B & \cdots & a_{mn}\,B\end{array}\right)
\end{equation}
In this representation the four components of the DKP spinor satisfy the following set of coupled equations:
\begin{multline}\label{a4}
\left(\begin{array}{c}
-2iH\eta\partial_\eta\varphi_1+iH\eta\widetilde{\partial}_1\varphi_2+iH\eta\widetilde{\partial}_1\varphi_3-iH\varphi_1-iH\varphi_4-m\varphi_1\\
-iH\eta\left(\widetilde{\partial}_1\varphi_1-\widetilde{\partial}_1\varphi_4\right)-m\varphi_2\\
-iH\eta\left(\widetilde{\partial}_1\varphi_1-\widetilde{\partial}_1\varphi_4\right)-m\varphi_3\\
2iH\eta\partial_\eta\varphi_4-iH\eta\widetilde{\partial}_1\varphi_2-iH\eta\widetilde{\partial}_1\varphi+iH\varphi_1+iH\varphi_4-m\varphi_4
\end{array}\right)+\frac{1}{2}\theta H
\left(\begin{array}{c}
-\widetilde{\partial}_1^2\varphi_2-\widetilde{\partial}_1^2\varphi_3\\
\widetilde{\partial}_1^2\varphi_1-\widetilde{\partial}_1^2\varphi_4\\
\widetilde{\partial}_1^2\varphi_1-\widetilde{\partial}_1^2\varphi_4\\
\widetilde{\partial}_1^2\varphi_2+\widetilde{\partial}_1^2\varphi_3
\end{array}\right)\\=\frac{i}{2}\theta H\left(\begin{array}{c}
-2iH\eta\partial_\eta\widetilde{\partial}_1\varphi_1+iH\eta\widetilde{\partial}_1^2\varphi_2+iH\eta\widetilde{\partial}_1^2\varphi_3-iH\widetilde{\partial}_1\varphi_1-iH\widetilde{\partial}_1\varphi_4-m\widetilde{\partial}_1\varphi_1\\
-iH\eta\left(\widetilde{\partial}_1^2\varphi_1-\widetilde{\partial}_1^2\varphi_4\right)-m\widetilde{\partial}_1\varphi_2\\
-iH\eta\left(\widetilde{\partial}_1^2\varphi_1-\widetilde{\partial}_1^2\varphi_4\right)-m\widetilde{\partial}_1\varphi_3\\
2iH\eta\partial_\eta\widetilde{\partial}_1\varphi_4-iH\eta\widetilde{\partial}_1^2\varphi_2-iH\eta\widetilde{\partial}_1^2\varphi+iH\widetilde{\partial}_1\varphi_1+iH\widetilde{\partial}_1\varphi_4-m\widetilde{\partial}_1\varphi_4
\end{array}\right).
\end{multline}

The plane-wave solutions $\varphi_i=e^{i\,k\,x}\,\widetilde{\varphi}_{i}$\,, with $i=\overline{1,4}$, imply that eq. \eqref{a4} is given by:
\begin{align}
\left[i\,H\,\eta\left(2+H\,\theta\,k\right)\partial_\eta+\frac{i\,\theta\,H\,k}{2}+i\,H+m\right]\widetilde{\varphi}_1&=-\left(H\,\eta\,k-\frac{1}{2}\,\theta\,H\,k^2+\theta\,H^2\,\frac{\eta}{2}\,k^2\right)
\left(\widetilde{\varphi}_2+\widetilde{\varphi}_3\right)\notag\\
&\phantom{=\,}-i\,H\left(1+\frac{\theta\,H\,k}{2}\right)\widetilde{\varphi}_4\,,\label{bb1}\\
m\left(1+\frac{1}{2}\,\theta\,H\,k\right)\widetilde{\varphi}_2&=\left(H\,\eta\,k-\frac{1}{2}\,\theta\,H\,k^2+\frac{1}{2}\,\theta\,H^2\,\eta\,k^2\right)
\left(\widetilde{\varphi}_1-\widetilde{\varphi}_4\right),\label{bb}\\
m\left(1+\frac{1}{2}\,\theta\,H\,k\right)\widetilde{\varphi}_3&=\left(H\,\eta\,k-\frac{1}{2}\,\theta\,H\,k^2+\frac{1}{2}\,\theta\,H^2\,\eta\,k^2\right)\left(\widetilde{\varphi}_1-\widetilde{\varphi}_4\right),\label{cc}\\
\left[i\,H\,\eta\,\left(2+H\,\theta\,k\right)\partial_\eta+i\,\frac{\theta\,H\,k}{2}+\left(i\,H-m\right)\right]\widetilde{\varphi}_4&=\left(H\,\eta\,k-\frac{1}{2}\,\theta\,H\,k^2+\theta\,H^2\,\frac{\eta}{2}\,k^2\right)
\left(\widetilde{\varphi}_2+\widetilde{\varphi}_3\right)+\notag\\
&\phantom{=\,}+iH\left(1+\frac{\theta\,H\,k}{2}\right)\widetilde{\varphi}_1\,.\label{cc1}
\end{align}
The two equations \eqref{bb} and \eqref{cc} imply that:
\begin{equation}
\widetilde{\varphi}_2=\widetilde{\varphi}_3\,.\label{eq1}
\end{equation}
Using \eqref{eq1}, the set of equations \eqref{bb1}-\eqref{cc1} become:
\begin{align}
\left(\partial_\eta+\frac{1}{2\,\eta}-\frac{i\,m}{2\,H\,\eta}\right)\widetilde{\varphi}_1+\frac{1}{2\,\eta}\,\widetilde{\varphi}_4-i\,k\left(1-\frac{\theta\,k}{2\,\eta}\right)\widetilde{\varphi}_2&=0\,,\label{K1}\\
\left(\partial_\eta+\frac{1}{2\,\eta}+\frac{i\,m}{2\,H\,\eta}\right)\widetilde{\varphi}_4+\frac{1}{2\,\eta}\,\widetilde{\varphi}_1-i\,k\left(1-\frac{\theta\,k}{2\,\eta}\right)\widetilde{\varphi}_2&=0\,,\label{K2}\\
H\,k\,\eta\left(1-\frac{1}{2}\,\frac{\theta\,k}{\eta}\right)\left(\widetilde{\varphi}_1-\widetilde{\varphi}_4\right)&=m\,\widetilde{\varphi}_2\,.\label{K3}
\end{align}
From \eqref{K1} and \eqref{K2} we write:
\begin{align}
\partial_\eta\left(\widetilde{\varphi}_1-\widetilde{\varphi}_4\right)&=\frac{i\,m}{2\,H\,\eta}\left(\widetilde{\varphi}_1+\widetilde{\varphi}_4\right),\label{K4}\\
\left(\partial_\eta+\frac{1}{\eta}\right)\left(\widetilde{\varphi}_1+\widetilde{\varphi}_4\right)-\frac{i\,m}{2\,H\,\eta}\left(\widetilde{\varphi}_1-\widetilde{\varphi}_4\right)&=2\,i\,k\left(1-\frac{\theta\,k}{2\,\eta}\right)
\widetilde{\varphi}_2\,.\label{K5}
\end{align}
Inserting eq. \eqref{K3} into \eqref{K5} yields:
\begin{equation}
\left(\partial_\eta+\frac{1}{\eta}\right)\left(\widetilde{\varphi}_1+\widetilde{\varphi}_4\right)=\left[\frac{i\,m}{2\,H\,\eta}+\frac{2\,H\,k^2\,\eta}{m}\left(1-\frac{\theta\,k}{\eta}\right)\right]
\left(\widetilde{\varphi}_1-\widetilde{\varphi}_4\right).\label{K6}
\end{equation}
Substituting \eqref{K4} into \eqref{K6} we obtain:
\begin{equation}
\left\{\partial_\eta^2+\frac{2}{\eta}\,\partial_\eta+\frac{m^2}{4\,H^2\,\eta^2}+k^2\left(1-\frac{\theta\,k}{\eta}\right)\right\}\left(\widetilde{\varphi}_1-\widetilde{\varphi}_4\right)=0\,.\label{K7}
\end{equation}
Making the transformations $\left(\widetilde{\varphi}_1-\widetilde{\varphi}_4\right)=\frac{1}{\eta}\,B\left(\eta\right)$ and $z=2\,i\,k\,\eta$, equation \eqref{K7} becomes:
\begin{equation}
\left[\partial_z^2-\frac{1}{4}+i\,\frac{\theta\,k^2}{2\,z}+\frac{\frac{1}{4}-\left(\frac{1}{4}-\frac{m^2}{4\,H^2}\right)}{z^2}\right]\widetilde{B}\left(z\right)=0.
\end{equation}
This equation has the well-known Whittaker solutions \cite{Abramowitz}:
\begin{equation}
\left[\partial_z^2-\frac{1}{4}+\frac{\lambda}{z}+\frac{\frac{1}{4}-\mu^2}{z^2}\right]W_{\lambda,\mu}\left(z\right)=0\,.
\end{equation}

Then
\begin{equation}\label{f1}
\begin{aligned}
\widetilde{\varphi}_1-\widetilde{\varphi}_4&=\frac{1}{\eta}\,W_{\lambda,\mu}\left(2\,i\,k\,\eta\right),\\
\lambda&=\frac{i\,\theta\,k^2}{2}\,,\\
\mu&=i\left\vert\mu\right\vert=i\left(\frac{m^2}{4\,H^2}-\frac{1}{4}\right)^{1/2}\,.
\end{aligned}
\end{equation}
From equation \eqref{K3} we have:
\begin{align}
\widetilde{\varphi}_2&=\widetilde{\varphi}_3=\frac{H\,k\,\eta}{m}\left(1-\frac{1}{2}\,\frac{\theta\,k}{\eta}\right)\left(\widetilde{\varphi}_1-\widetilde{\varphi}_4\right)\notag\\
&=\frac{H\,k}{m}\left(1-\frac{1}{2}\,\frac{\theta\,k}{\eta}\right)W_{\lambda,\mu}\left(2\,i\,k\,\eta\right),\label{K00}
\end{align}
and \eqref{K4} gives:
\begin{equation}
\widetilde{\varphi}_1+\widetilde{\varphi}_4=\frac{2\,i\,H}{m\,\eta}\,W_{\lambda,\mu}\left(2\,i\,k\,\eta\right)-\frac{2\,i\,H}{m}\,\partial_\eta\,W_{\lambda,\mu}\left(2\,i\,k\,\eta\right).\label{f2}
\end{equation}

Using the property \cite{Abramowitz}:
\begin{equation}
z\,\partial_z\,W_{\lambda,\mu}\left(z\right)=\left(\frac{1}{2}\,z-\lambda\right)W_{\lambda,\mu}\left(z\right)-W_{\lambda+1,\mu}\left(z\right),
\end{equation}
and from eqs. \eqref{f1}, \eqref{K00} and \eqref{f2}, we obtain the general solution of the non-commutative DKP equation \eqref{eq:fourteen} as follows:
\begin{equation}
\left(\begin{array}{c}
\widetilde{\varphi}_1\\
\widetilde{\varphi}_2\\
\widetilde{\varphi}_3\\
\widetilde{\varphi}_4
\end{array}\right)=\left(\begin{array}{c}
\frac{1}{2\eta}\left[\left(1+\frac{2iH}{m}\right)W_{\lambda,\mu}\left(2\,i\,k\,\eta\right)-\frac{2\,i\,H}{m}\left(\left(i\,k\,\eta-\lambda\right)
W_{\lambda,\mu}\left(2\,i\,k\,\eta\right)-W_{\lambda +1,\mu}\left(2\,i\,k\,\eta\right)\right)\right]\\
\frac{H\,k}{m}\left(1-\frac{1}{2}\,\frac{\theta\,k}{\eta}\right)W_{\lambda,\mu}\left(2\,i\,k\,\eta\right)\\
\frac{H\,k}{m}\left(1-\frac{1}{2}\,\frac{\theta\,k}{\eta}\right)W_{\lambda,\mu}\left(2\,i\,k\,\eta\right)\\
\frac{1}{2\,\eta}\left[\left(-1+\frac{2iH}{m}\right)W_{\lambda,\mu}\left(2\,i\,k\,\eta\right)-\frac{2\,i\,H}{m}\left(\left(i\,k\,\eta-\lambda\right)
W_{\lambda,\mu}\left(2ik\eta\right)-W_{\lambda+1\mu}\left(2ik\eta\right)\right)\right]
\end{array}\right)\notag
\end{equation}

In order to construct the positive and negative frequency modes, we use the asymptotic behavior of the solutions $\psi $ and compare the result with that obtained by solving the Hamilton-Jacobi relativistic equation. In fact,
for $z\rightarrow 0$, one can show that the positive and negative frequency solutions, $\psi^+(z\rightarrow 0)$ and $\psi^-(z\rightarrow 0)$, are respectively given by the following asymptotic forms:
\begin{equation}
\psi^+(z\rightarrow 0)\approx C_0^+\,M_{\lambda,\mu}\left(z\right),
\end{equation}
and
\begin{equation}
\psi^-(z\rightarrow 0)\approx C_0^+\left(-1\right)^{-\mu+1/2}\,M_{\lambda,-\mu}\left(z\right),
\end{equation}
where the Whittaker function $M_{\lambda,\mu}\left(z\right)$ has the asymptotic behaviour:
\begin{equation*}
M_{\lambda,\mu}\left(z\right)\approx e^{-z/2}\,z^{\mu+1/2}\,,\qquad z\ll 1\,,
\end{equation*}
and $C_0^+$ is a normalization function. Similarly, for $z\rightarrow \infty$, the corresponding positive and negative frequency modes are:
\begin{equation}
\psi^+(z\rightarrow \infty)\approx C_{\infty}^{+}\,W_{\lambda,\mu}\left(z\right),\label{aa}
\end{equation}
and
\begin{equation}
\psi^-(z\rightarrow \infty)\approx C_\infty^-\,W_{-\lambda,\mu}\left(z\right),\label{aa1}
\end{equation}
where $C_\infty^\pm$ are normalization functions. Now, using the fact that \cite{Abramowitz}:
\begin{equation}
W_{\lambda,\mu}\left(z\right)=\frac{\Gamma\left(-2\,\mu\right)}{\Gamma\left(\frac{1}{2}-\mu-\lambda\right)}\,M_{\lambda,\mu}\left(z\right)+\frac{\Gamma\left(2\,\mu\right)}{\Gamma\left(\frac{1}{2}+\mu-\lambda\right)}\, M_{\lambda,-\mu}\left(z\right),
\end{equation}
where $\Gamma\left(z\right)$ is the Euler Gamma function, and
\begin{align}
\left(W_{-\lambda,\mu}\left(-z\right)\right)^\ast&=W_{\lambda,\mu}\left(z\right),\\
\left(M_{\lambda,\mu}\left(z\right)\right)^\ast&=M_{\lambda,-\mu}\left(z\right),
\end{align}
we can deduce that:
\begin{equation}
\psi^+(z\rightarrow 0)=\frac{\Gamma\left(-2\,\mu\right)}{\Gamma\left(\frac{1}{2}-\mu-\lambda\right)}\,\psi^+(z\rightarrow\infty)+\frac{\Gamma\left(2\,\mu\right)}{\Gamma\left(\frac{1}{2}+\mu-\lambda\right)}\,
e^{i\,\pi\left(\mu-1/2\right)}\,\psi^-(z\rightarrow\infty)\,.
\end{equation}

Now, since we are able to obtain the single-particle states in the vicinity of $z\rightarrow 0$ and $z\rightarrow\infty $, we can then compute the density of particles created by the non-commutative curved space-time and
electromagnetic field. In fact, with the help of the Bogouliugov transformation, the positive frequency mode at $z\rightarrow 0$ can be written as a linear combination of the positive and negative frequency modes at $z\rightarrow\infty$ in the form:
\begin{equation}
\psi^+(z\rightarrow 0)=\alpha\,\psi^+(z\rightarrow \infty)+\beta\,\psi^-(z\rightarrow \infty)\,,\label{d1}
\end{equation}
where we used eqs. \eqref{aa}-\eqref{aa1}, as well as the normalization condition:
\begin{equation}
\left\vert \alpha \right\vert^2-\left\vert\beta\right\vert^2=1\,,\label{d2}
\end{equation}
where the operators $\alpha$ and $\beta$ are the Bogoliubov coefficients.

We now turn to the non-commutative density $\widehat{n}$ of particles created by the non-commutative curved space-time. For this we use eqs. \eqref{d1} and \eqref{d2} so as to arrive at:
\begin{equation}
\widehat{n}=\left[\left(\frac{\left\vert\beta\right\vert^2}{\left\vert\alpha\right\vert^2}\right)^{-1}-1\right]^{-1}\,.\label{d3}
\end{equation}
Using the relation:
\begin{equation}
\Gamma\left(i\,z+1/2\right)=\frac{\pi}{\cosh\pi z}\,,
\end{equation}
we obtain after direct simplifications:
\begin{equation}
\frac{\left\vert\beta\right\vert^2}{\left\vert\alpha\right\vert^2}=\frac{\cosh\pi\left\vert-\mu+\lambda\right\vert}{\cosh\pi\left\vert\mu+\lambda\right\vert}\,e^{-2\,\pi\left\vert\mu\right\vert}\,.
\end{equation}
Then the non-commutative number density defined in eq. \eqref{d3} can be written as:
\begin{equation}
\widehat{n}=\frac{1}{e^{2\,\pi\left\vert\mu\right\vert}\,f\left(k\right)-1}\,,
\end{equation}
where
\begin{equation}
f\left(k\right)=\frac{\cosh\pi\left\vert-\mu+\lambda\right\vert}{\cosh\pi\left\vert\mu+\lambda\right\vert}\,.
\end{equation}
We have thus obtained a Bose-Einstein-like distribution. It is also very important to consider the massive case and see the behaviour of the number density, and derive some of the related thermodynamical quantities. Taking
$m/H\gg 1$, it is easy to show that the probability $f\left(k\right)$ takes the form:
\begin{equation*}
f\left(k\right)\approx\exp\left(-\pi\,\theta\,k^2\right).
\end{equation*}
Then the non-commutative number density $\widehat{n}$ can be written up to second order in the non-commutativity parameter as:
\begin{equation}
\widehat{n}=\frac{1}{e^{2\,\pi\left(\frac{m}{2\,H}-\frac{\theta\,k^2}{2}\right)}-1}\,.\label{M}
\end{equation}

The spectrum of created particles is thermal Bose-Einstein distribution with the following non-commutative temperature:
\begin{align}\label{eq:sixty}
\widetilde{T}&=\frac{H}{2\,\pi}+\frac{H^2\,\theta}{2\,\pi\,m}\,k^2\notag\\
&\approx T_0\,\exp\left(\frac{H}{m}\,\theta\,k^2\right),
\end{align}
where $T_0=\frac{H}{2\,\pi}$ is the ordinary Hawking temperature in commutative de-Setter space \cite{8a}. It is worthwhile to recall that the non-commutative temperature correction in eq. \eqref{eq:sixty} is directly related to the micro-canonical mean kinetic energy per degree of freedom, by virtue of the equipartition theorem:
\begin{equation*}
k_B\,T=\frac{\hbar^2\,k^2}{m}\,.
\end{equation*}
This shows how particle production occurs under the effects of non-commutativity, which leads us to a better understanding of Hawking radiation from black holes.

\begin{figure}[th]
\centering
\includegraphics[width=0.49\textwidth]{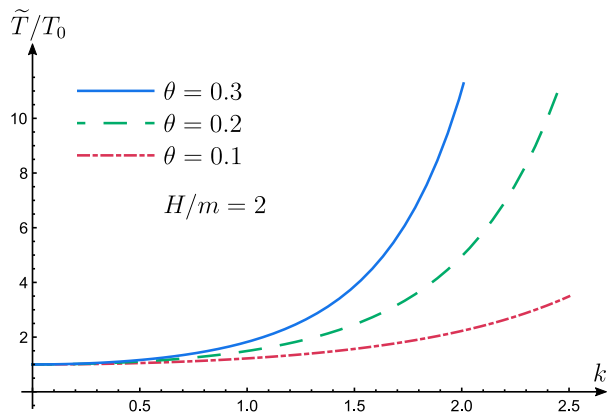}
\caption{Evolution of $\widetilde{T}/T_{0}$ in non-commutative curved space-time.}
\label{fig1}
\end{figure}

Figure \ref{fig1} shows the evolution of $\widetilde{T}/T_{0}$ in non-commutative curved space-time as a function of $k$ for various values of $\theta$ ($H/m=2$).

We note that in figure \ref{fig1} the fluctuations which increase the temperature, proportional to the non-commutativity parameter, lead to a decreasing non-commutative effective cosmological constant. Thus, the generalized second law implies that the de-Sitter space is unstable to fluctuations which decrease the non-commutativity effects. Here we note that the expansion of the universe is therefore the result of
increasing its temperature, which provides the energy required for this expansion.

\begin{figure}[ht]
\centering
\includegraphics[width=0.49\textwidth]{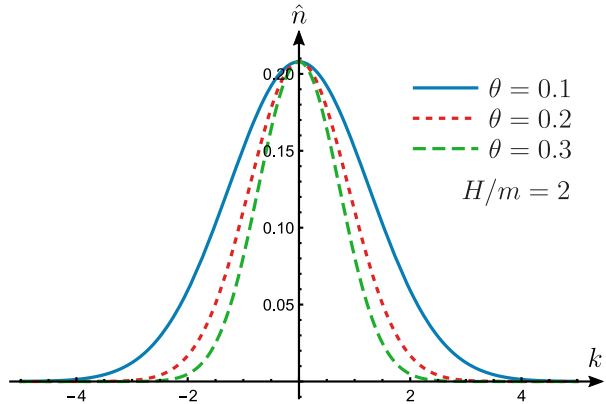}
\caption{Non-commutative number density $\widehat{n}$ as a function of $k$ for various values of $\theta$ ($H/m=2$).}
\label{fig2}
\end{figure}

We can take the expression \eqref{M} as:
\begin{equation*}
\widehat{n}\left(k\right)\approx\exp\left[-\frac{m\,\pi}{H}\left(1+\frac{H\,\theta\,k^2}{m}\right)\right]
\end{equation*}
Here we clearly demonstrated that the effect of the non-commutativity plays the role of an electric and a magnetic field. It is reassuring to note that in the limit $\theta =0$, the commutative result is recovered (see ref. \cite{12a}):
\begin{equation}
\widehat{n}=n_0=\exp\left(-\frac{m\,\pi}{H}\right).
\end{equation}
In figure \ref{fig2}, we plot the non-commutative number density $\widehat{n}$ as a function of the variable $k$ for various values of $\theta$ ($H/m=2$). Notice that if $\theta$ increases, $\widehat{n}(k)$ decreases.

\begin{figure}[ht]
\centering
\includegraphics[width=0.49\textwidth]{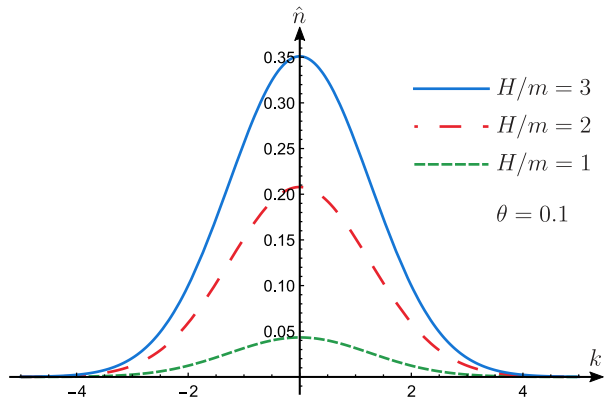}
\caption{Non-commutative number density $\widehat{n}$ as a function of $k$ for various values of $H/m$ ($\theta=0.1$).}
\label{fig3}
\end{figure}

Figure \ref{fig3} represents the non-commutative number density $\widehat{n}$ as a function of the variable $k$ for various values of $H/m$ with $\theta=0.1$.

\begin{figure}[ht]
\centering
\includegraphics[width=0.49\textwidth]{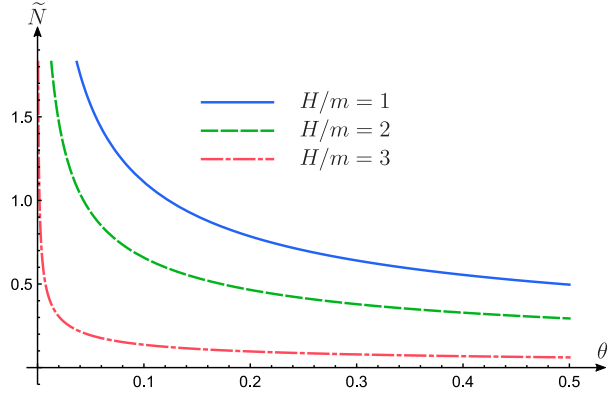}
\caption{Dependence of $\widetilde{N}$ on $\protect\theta$ for various values of $H/m$.}
\label{fig4}
\end{figure}

Comparing figures \ref{fig1} and \ref{fig3} one can easily see that the non-commutativity effects play the dominant role compared to those of gravitational fields. For nonzero $\theta$, i.e. in the non-commutative case, the total number is given by:
\begin{equation}
\widetilde{N}=\int_{-\infty}^\infty dk\,n\left(k\right)=\frac{\exp\left(-\pi\,m/H\right)}{\sqrt{\theta}}\,.
\end{equation}
The dependence of $\widetilde{N}$ on the intensity of the non-commutative parameter $\theta$ in curved space-time for various values of $H/m$ is shown in figure \ref{fig4}.
Notice that the non-commutative number of created particles increases due to the weaker values of the non-commutativity parameter.

\section{Conclusion}

In this work we used the Seiberg-Witten maps and the Moyal product up to first order in the non-commutativity parameter and modified the DKP equation in vacuum for a two-dimensional non-commutative curved space-time. After solving this equation we calculated the number density of created particles by applying the Bogoliubov transformations and the quasi-classical limit for identifying the positive and negative frequency modes. Consequently we showed that the non-commutativity plays the same role as the electric and magnetic fields or gravity for the particle creation process. The non-commutative density of created particles is found to decrease as $1/\sqrt{\theta}\sim\Lambda_{\mathrm{NC}}$, so that the rate of particle creation increases whenever a non-commutativity parameter is small, and this corresponds to the spirit of quantum mechanics. We can say that the non-commutativity parameter $\theta$ determines the fundamental cell concretization much in the same way as the Planck constant $h$ discretizes the phase space.


\begin{thebibliography}{99}
\bibitem{1a} R. Casana, V.Ya. Fainberg, B.M. Pimentel, J.S. Valverde,
Phys.Lett. A316 (2003) 33

\bibitem{2a} R. Casana, B.M. Pimentel, J.T. Lunardi and R.G. Teixeira,
Int.J.Mod.Phys. A17 (2002) 4197

\bibitem{3a} R. Casana, V.Y. Fainberg, J.T. Lunardi, B.M. Pimentel and R.G.
Teixeira, Class. Quant. Grav. 20 (2003) 2457

\bibitem{4a} R. Casana, J.T. Lunardi, B.M. Pimentel and R.G. Teixeira,
Gen.Rel.Grav. 34 (2002) 1941-1951

\bibitem{5a} S. Haouat and R. Chekireb, Eur.Phys.J. C72 (2012) 2034

\bibitem{6a} S. Haouat and R. Chekireb, Mod.Phys.Lett. A26 (2011) 2639

\bibitem{7a} Sang Pyo Kim, Grav.Cosmol. 20 (2014) 193

\bibitem{8a} E. Ersin Kangal, Hilmi Yanar, Ali Havare, Kenan Sogut, Ann.
Phys. 343 (2014) 40

\bibitem{9a} Garriga, J. Phys.Rev. D49 (1994) 6343

\bibitem{10a} Havare, Ali et al. Nucl.Phys. B682 (2004) 457

\bibitem{11a} Haro, Jaume et al. J.Phys. A41 (2008) 372003

\bibitem{12a} Biswas, S. et al. Class.Quant.Grav. 12 (1995) 1591

\bibitem{13a} Biswas, S. et al. Gen.Rel.Grav. 34 (2002) 665-678

\bibitem{14a} Villalba, Victor M. Phys.Rev. D52 (1995) 3742

\bibitem{15a} N. Seiberg and E.Witten, JHEP 032 (1999) 9909.

\bibitem{16a} L. Khodja and S. Zaim, Int. J. Mod. Phys. A27 (2012) 1250100.

\bibitem{17a} N.Mebarki,S.Zaim,L.Khodja and H.Aisaoui, Phys. Scripta 78
(2008) 045101.

\bibitem{18a} M. Chaichian, A. Tureanu and G. Zet, Phys. Lett. B
660 (2008) 573

\bibitem{19a} S. Fabi, B. Harms and A. Stern, Phys.Rev. D78 (2008) 065037

\bibitem{cronecker} A. J. Laub, Matrix Analysis for Scientists and
Engineers, SIAM, PA, 2005 (chapter 13).

\bibitem{Abramowitz} M. Abramowitz, I. Stegun, Handbook of Mathematical
Functions, Dover, New York, 1974.
\end{thebibliography}
\end{document}